\documentclass[twocolumn,twoside,preprintnumbers,amsmath,amssymb,showkeys]{revtex4}

\usepackage{epsfig}

\usepackage{graphicx}

\usepackage{fancyhdr}
\usepackage{pslatex}

\newcommand {\Alm} {{\mbox{$A_{l,m}$}}}
\newcommand {\AlmQ} {{\mbox{$A_{l,m}(Q)$}}}

\newcommand {\genbod} {{\sc genbod}~}

\begin{document}
\title {Global Conservation Laws and Femtoscopy of Small Systems}

\author{Zbigniew Chaj\c{e}cki$^1$ and Mike Lisa$^1$}

\affiliation{$^1$ Department of Physics, Ohio State University,
1040 Physics Research Building, 191 West Woodruff Ave, Columbus, OH 43210, USA}

\begin{abstract}
It is increasingly important to understand, in detail, two-pion correlations
measured in $p+p$ and $d+A$ collisions.  In particular, one wishes to understand
the femtoscopic correlations, in order to compare to similar measurements in heavy
ion collisions.  However, in the low-multiplicity final states of these systems,
global conservation laws generate significant $N$-body correlations which project
onto the two-pion space in non-trivial ways and complicate the femtoscopic analysis.
We discuss a model-independent formalism to calculate and account for these correlations
in measurements.
\end{abstract}

\vskip -1.35cm
\maketitle
\thispagestyle{fancy}
\setcounter{page}{1}
\bigskip

\section{INTRODUCTION}

The unique and distinguishing feature of
heavy ions is their large (relative to the confinement scale) size and the possibility to generate
{\it bulk} systems which may be described in thermodynamic terms, allowing to discuss the Equation
of State of strongly-interacting {\it matter}.  The primary evidence for the creation of bulk 
matter at the highest energies is the existence of strong collective flow~\cite{Ollitrault:1992bk}.
The dominant feature of flow is the correlations between space and momentum which it generates; thus,
momentum-only observables such as $p_T$ spectra and azimuthal 
anisotropies~\cite{Adams:2005dq,Adcox:2004mh,Back:2004je,Arsene:2004fa} represent only an indirect
projection of the effect.  Femtoscopic measurements access space as a function of particle momentum,
thus providing the most direct probe of the most crucial feature of heavy ion collisions~\citep[c.f. e.g.][]{Lisa:2005dd}.
In particular, flow is manifest by a negative correlation between the ``HBT radius'' and the transverse
mass ($m_T$) of the particles~\cite{Pratt:1984su}.

Clearly, then, a detailed understanding of femtoscopic measurements in heavy ion collisions
is crucial to proving the existence of, or probing the nature of, the bulk system generated
in the collision.  It is in fact possible to quantitatively interpret both
the femtoscopic and momentum-only observations at RHIC in consistent,
flow-dominated models of the system~\citep[e.g.][]{Retiere:2003kf}. 
All seems well.

However, two-pion femtoscopic measurements are also common in $e^++e^-$ or $p+p(\bar{p})$
collisions~\cite{Alexander:2003ug}.  In these collisions, too, ``HBT radii'' are observed
to fall with $m_T$.  Speculations of the physics behind this observation have included
Heisenberg uncertainty-based arguments, string-breaking phenomena, and temperature gradients;
an excellent overview may be found in~\cite{Kittel:2005fu}.
Typically, however, one might not expect the system created in a $p+p$ collision to exhibit
{\it bulk} behaviour similar to that from heavy ion collisions.

{\it Quantitative} comparisons between femtoscopic measurements in $A+A$ and $p+p$ systems
have been complicated because techniques for event-mixing, frame definitions, and the like,
have been different in the particle-physics and heavy-ion communitites.  As importantly,
kinematic acceptance and collision energies are usually quite different.  
Recently, however, the STAR experiment has reported the first preliminary study of directly-comparable femtoscopic
measurements from $A+A$ and $p+p$ systems~\cite{Chajecki:2005zw} at the same $\sqrt{s_{NN}}$,
using the same detector, and with identical techniques.
The results indicate that the femtoscopic probe of flow-- falling ``HBT radii'' with $m_T$-- is
essentially {\it identical} in the small and large systems.  This might signal an unexpected
``universality'' in the spatial substructure of hadronic and heavy ion collisions.
Unravelling the physics behind this similarity might provide new insight into the dynamical
space-time substructure of {\it both} hadronic and heavy ion collisions.


Before drawing strong physics conclusions from ``HBT radii'' coming from fits to
the pion correlations measured in $p+p$ collisions, however, the measured correlation
functions {\it themselves} must be understood in detail.  The STAR data show clear
non-femtoscopic correlations which must be disentangled from the femtoscopic
ones~\cite{Chajecki:2005zw}.
Femtoscopic correlations are those which depend directly on the two-particle separation
distribution~\citep[c.f.][]{Lisa:2005dd}.  Non-femtoscopic correlations may arise from string
or jet fragmentation, resonance decay, or global conservation laws.

In this work, we explore the projection of $N$-body Energy and Momentum Conservation-Induced Correlations
(EMCICs) onto a two-particle relative momentum correlation function.
In Section~\ref{sec:SHD} we briefly discuss the harmonic representation
of the correlations which best illustrates the effect.  In Section~\ref{sec:GenBod} we discuss
EMCICs generated by a Monte Carlo event generator containing only global conservation laws.
A method to calculate analytically (but using distributions from the data) EMCICs is shown
in Section~\ref{sec:JYO}.  This provides an ``experimentalist's formula,'' given in Section~\ref{sec:formula},
useful to disentangle
EMCICs from the data, allowing a femtoscopic analysis of the correlation functions.  We summarize
in Section~\ref{sec:summary}.

\section{SPHERICAL HARMONIC DECOMPOSITION OF CORRELATION FUNCTIONS}
\label{sec:SHD}

\begin{figure}[t]
{\centerline{\includegraphics[width=0.3\textwidth]{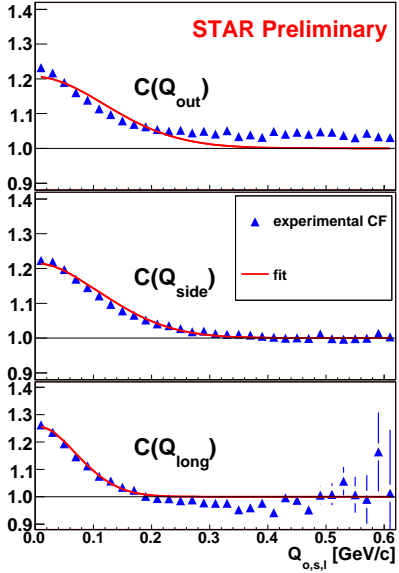}}}
\caption{Preliminary STAR two-pion correlation functions~\cite{Chajecki:2005zw} presented
as 1D projections in the Bertsch-Pratt decomposition.\label{fig:CartesianDAu}}
\end{figure}

At asymptotically high relative momentum $|\vec{q}|$ (or $|\vec{k^*}|$), femtoscopic contributions
to the the correlation function (those described by the Koonin-Pratt equation~\citep[discussed in][]{Lisa:2005dd})
must approach a constant value, usually normalized to unity, independent of the direction of $\vec{q}$.
Preliminary STAR measurements~\cite{Chajecki:2005zw} of small systems, Figure~\ref{fig:CartesianDAu},
show clear non-femtoscopic correlations in addition.  Also shown is a fit with the commonly Gaussian (with
Coulomb suppression) functional form~\cite{Lisa:2005dd}.  Clearly, the fit is a poor representation of the
data.  We stress, however, that it is not the (non-)Gaussian nature of the source at issue here; {\it any}
source function will lead to vanishing femtoscopic correlations at large $|\vec{q}|$ and will thus contradict
the data.

We further stress that the problem is not one of normalization.  Shown in the Figure is the common representation
of the 3-dimensional correlation function into three 1-dimensional axes~\cite[cf][]{Lisa:2005dd}.
The projections, then, are not independent and cannot be independently normalized.  The problem is
that the value approached at large $|\vec{q}|$ depends on the direction in $\vec{q}$ space.

One-dimensional projections present a limited
tool for exploring detailed structure of the three-dimensional correlation function.
The spherical harmonic decomposition (SHD) is a much more efficient representation of the
data which uses {\it all} of the data to show the shape of the correlation function in
3D $\vec{q}$ space.
There, the spherical harmonic coefficients $A_{l,m}$, which depend
on $Q\equiv |\vec{q}|$, are calculated as
%
\begin{eqnarray}
\label{eq:Alm}
\AlmQ = \sum_{{\rm bins} \thinspace i} C\left(Q,\cos\theta_i,\phi_i \right) \cdot \qquad\qquad\qquad\qquad\nonumber \\
         Y_{l,m}\left(\cos\theta_i,\phi_i\right) F_{l,m}\left(\cos\theta_i,\Delta_{\cos\theta},\Delta_\phi\right) ,
\end{eqnarray}
where $F_{l,m}$ represents a numerical factor correcting for finite bin sizes $\Delta_{\cos\theta}$ and $\Delta_\phi$;
it turns out not to depend on $\phi_i$.
The angles $\theta$ and $\phi$ are related to the Bertsch-Pratt Cartesian coordinate system through
\begin{equation}
q_{o} = Q\sin{\theta}\cos{\phi} , \quad
q_{s} = Q\sin{\theta}\sin{\phi} , \quad
q_{l} = Q\cos{\theta} .
\end{equation}
See~\cite{Chajecki:2005qm} for a complete discussion.

\begin{figure}[t!]
{\centerline{\includegraphics[width=0.45\textwidth]{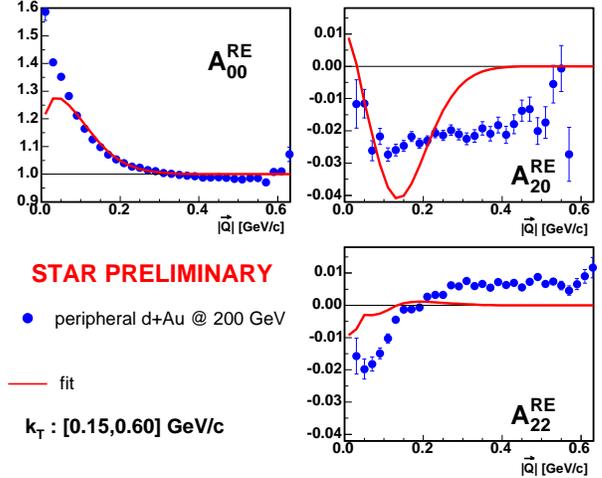}}}
\caption{Preliminary STAR two-pion correlation functions~\cite{Chajecki:2005zw} presented
in the SHD representation.\label{fig:SHDdAu}}
\end{figure}

Preliminary STAR correlation functions in the SHD representation~\cite{Chajecki:2005zw}
are shown in Figure~\ref{fig:SHDdAu}.
Coefficients for $l\geq 4$ are much less significant, compared to errorbars;
to good approximation, the non-femtoscopic behaviour is quadrupole ($l=2$) in nature.

The presence of non-femtoscopic correlations is clear from the non-vanishing behaviour of $A_{l\neq 0,m}$'s at large $Q$.
However, it is by no means clear that these contributions to the correlation function are {\it confined}
to large $Q$.  Thus, one cannot attempt to interpret the low-$Q$ region only in terms of femtoscopic correlations,
while parameterizing or ignoring the large-$Q$ region; see~\cite{Chajecki:2006sf} for further discussion.

\section{EMCICs GENERATED BY THE {\sc genbod} MONTE CARLO GENERATOR}
\label{sec:GenBod}

Non-femtoscopic correlations may arise from a variety of sources.
Jets will clearly induce momentum-space correlations between its fragmentation products.
While this cannot be discounted, the low momentum of the pions under consideration ($p_T \sim 0.2$~GeV)
puts us squarely in the region in which factorization breaks down and the jet interpretation
becomes significantly murkier.  We do not explore this possibility here.
In the kinematic region under consideration, string fragmentation may play a role; this is an
area for future study, though significant model-dependence will be present.

Resonances induce correlations among daughters as well; while these might even dominate $\pi^+\pi^-$
correlations, they should be negligible for identical pion correlations.  Collective bulk flow (e.g.
anisotropic elliptic flow) will generate $N$-body correlations which will project onto the two-body
space.  Non-femtoscopic correlations of the type observed by STAR in small systems are not, however,
observed in Au+Au collisions, despite the fact that elliptic flow is much larger there; therefore, we
do not believe that collective flow generates the observed effects.

Without doubt, one physical effect which {\it must} be at play is momentum and energy conservation.
As global conservation laws, these provide an $N$-body constraint on the event, which projects down
onto 2-body spaces.  The observed non-femtoscopic effects~\cite{Chajecki:2005zw} become more and
more significant as the multiplicity ($N$) of the event decreases, as expected from conservation laws.
It is these EMCICs which we focus on here.

To clearly understand the role of EMCICs, we would like to have events in which there is no other physics
involved besides the conservation laws.  Such a tool has been provided almost 40 years ago in the form of
the \genbod computer program~\cite[see ][ for an excellent writeup of the method and physics]{James:1968gu}
in the cern library.  
Given a requested total multiplicity ($N$), a list of masses ($m_i$) of emitted particles, and a
total amount of energy ($E_{\rm tot}$) to distribute among them, \genbod returns an event of random
momenta (four vectors $p_j$), subject only to the condition of energy and momentum conservation.  More importantly, it returns,
for each event, a weight proportional to the probability that the event will actually occur in nature.
This weight is proportional to the phasespace integral $R_N$
\begin{equation}
R_N = \int^{4N}\delta^4\left(P-\sum_{j=1}^N p_j \right)\prod_{i=1}^N\delta\left(p_i^2-m_i^2\right)d^4p_i ,
\end{equation}
where $P = \left(E_{\rm tot},\vec{0}\right)$ is the total momentum four-vector of the event.
See~\cite{James:1968gu} for a practical iterative prescription to calculate $R_N$.
Thus, it is a much different tool than, say RQMD, in which each event returned may be treated as equally
probable.

%

\begin{figure}[t]
{\centerline{\includegraphics[width=0.45\textwidth]{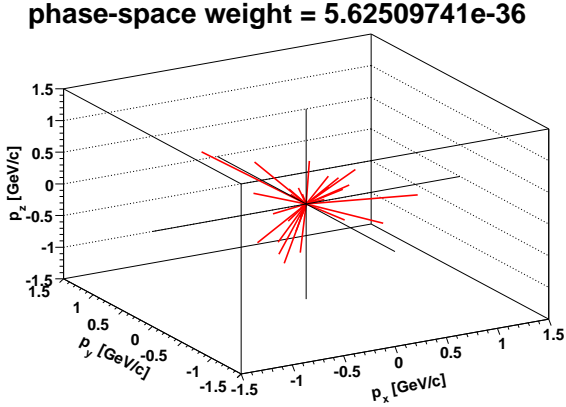}}}
\caption{A high-probability multiplicity-30 event calculated by \genbod.
Lines correspond to particle momenta $p_x,p_y,p_z$.\label{fig:LikelyEvent}}
\end{figure}

\begin{figure}[t]
{\centerline{\includegraphics[width=0.45\textwidth]{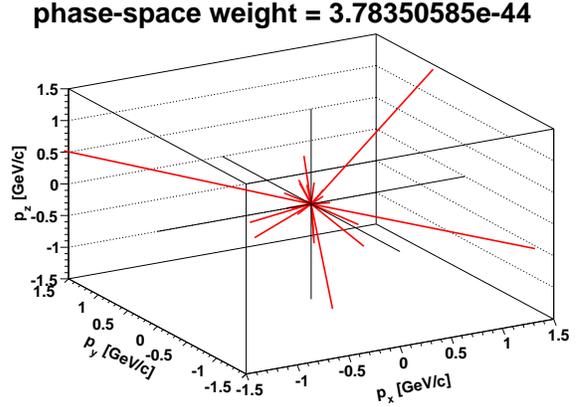}}}
\caption{A low-probability multiplicity-30 event calculated by \genbod.
Lines correspond to particle momenta $p_x,p_y,p_z$.\label{fig:UnlikelyEvent}}
\end{figure}

We select (via monte-carlo) \genbod events according to their weight and run them through identical software
as used for experimental analysis.  Fortunately, the code is fast, since one must calculate large statistics
from which to select.  This is because the phase-space weights vary by large factors.  As a very extreme case,
Figures~\ref{fig:LikelyEvent} and~\ref{fig:UnlikelyEvent} show a likely and unlikely event, respectively,
for multiplicity $N=30$.  As one would expect, the ``rounder'' event is more likely, though one might be surprised
by the factor of a hundred million between the probabilities.

\begin{figure}[t]
{\centerline{\includegraphics[width=0.45\textwidth]{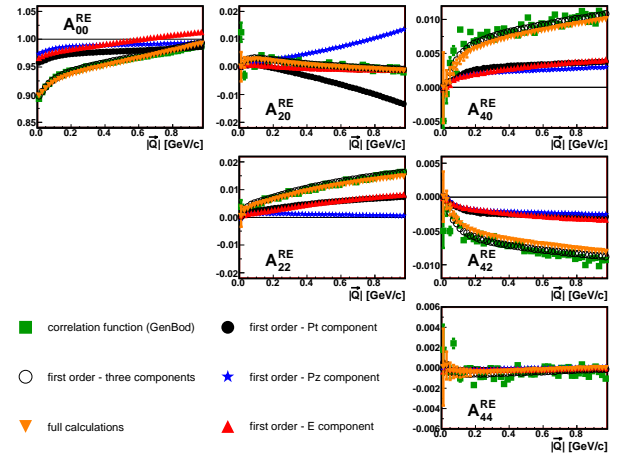}}}
\caption{SHD coefficients for \genbod-generated events consisting of 18 pions, as measured in the pair CMS frame.
Green squares are \Alm's from the \genbod events.  For discussion of the
other symbols, see Section~\ref{sec:JYO}.\label{fig:GenbodMult18-noEtaCut}}
\end{figure}

\begin{figure}[t]
{\centerline{\includegraphics[width=0.45\textwidth]{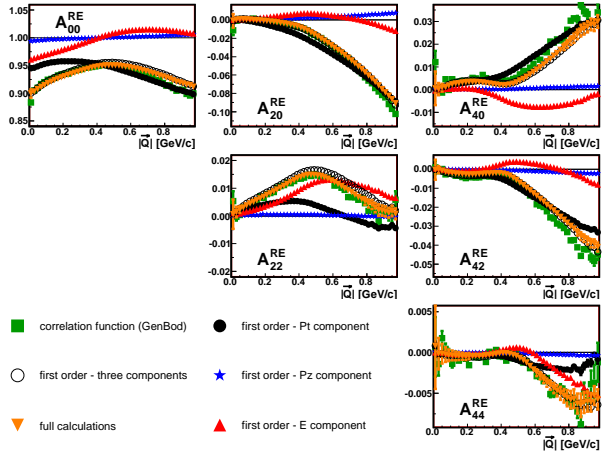}}}
\caption{Same as in Figure~\ref{fig:GenbodMult18-noEtaCut}, except only using pions with
$|\eta|<0.5$ in the correlation function.\label{fig:GenbodMult18-withEtaCut}}
\end{figure}

Figures~\ref{fig:GenbodMult18-noEtaCut} and~\ref{fig:GenbodMult18-withEtaCut}
show the \Alm's calculated by \genbod
for 18-pion events without and with a selection of $|\eta|<0.5$, respectively.
Note that this cut applies to the pions which are used in the analysis, {\it not}
to the set of particles for which energy and momentum is conserved; energy and momentum
is always conserved for the full event.
Clearly visible are significant and  nontrivial \Alm's 
due only to EMCICs.  We observe also that the $l=4$ coefficients are about an order of
magnitude smaller than the $l=2$ ones; this is generically expected~\citep[cf ][]{Chajecki:2005qm}.
Comparing the two figures, it is clear that kinematical selection has significant effect on
the EMCIC effects.  Also significant (but not shown) is whether one includes other species
(say protons) into the mix of emitted particles.


\begin{figure}[t]
{\centerline{\includegraphics[width=0.45\textwidth]{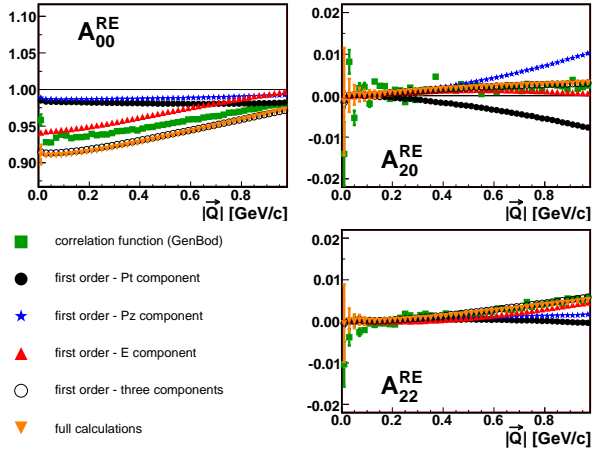}}}
\caption{SHD coefficients for \genbod-generated events consisting of 18 pions, as measured in the pair LCMS frame.
Green squares  are \Alm's from the \genbod events.  For discussion of the
other symbols, see Section~\ref{sec:JYO}.\label{fig:GenbodMult18LCMSK0p5}}
\end{figure}

\begin{figure}[t]
{\centerline{\includegraphics[width=0.45\textwidth]{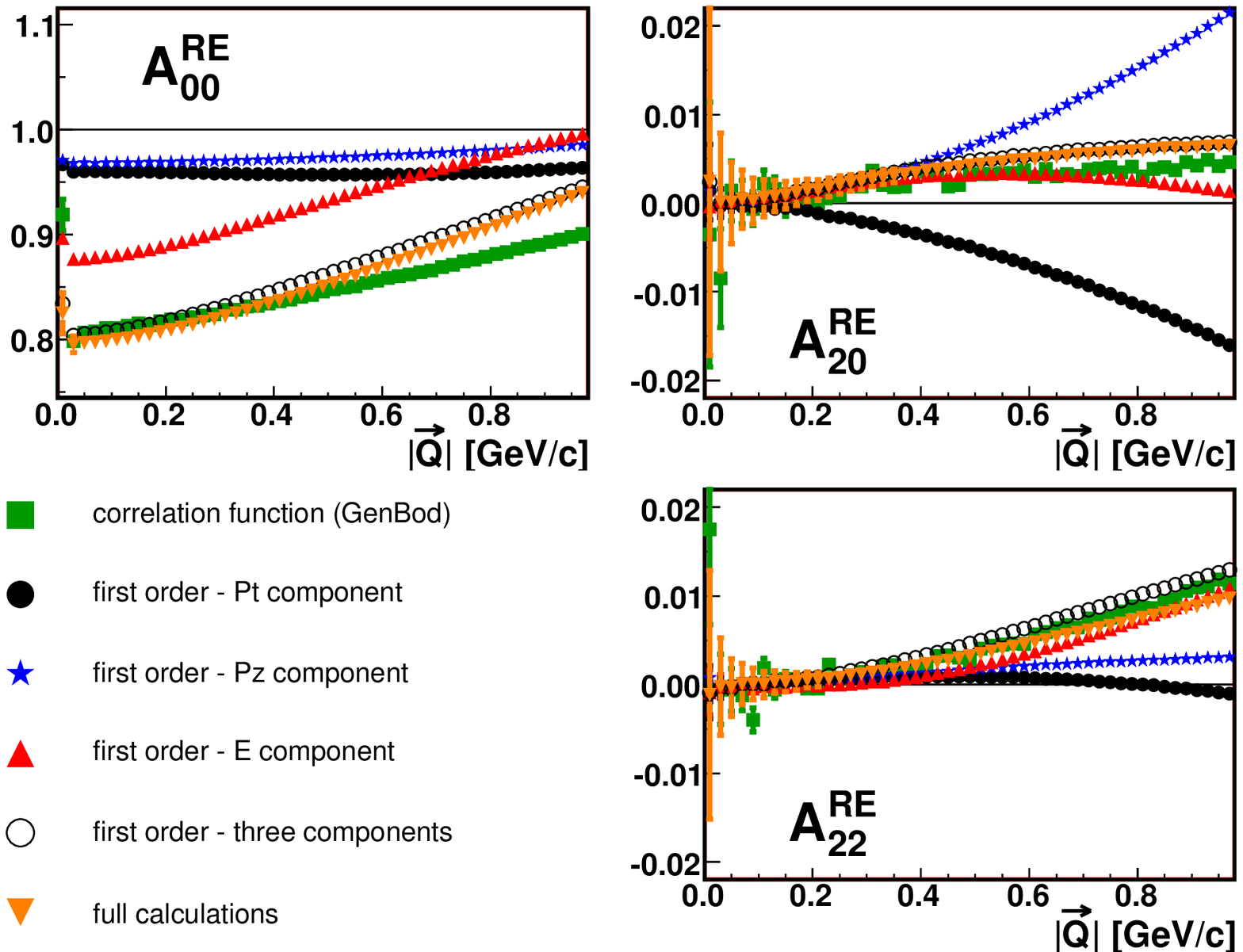}}}
\caption{Same as Figure~\ref{fig:GenbodMult18LCMSK0p5}, but for 9-pion events.\label{fig:GenbodMult9LCMSK0p5}}
\end{figure}

\begin{figure}[t]
{\centerline{\includegraphics[width=0.45\textwidth]{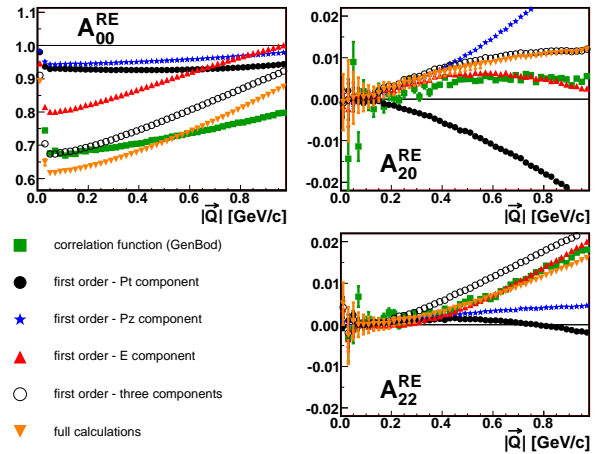}}}
\caption{Same as Figure~\ref{fig:GenbodMult18LCMSK0p5}, but for 6-pion events.\label{fig:GenbodMult6LCMSK0p5}}
\end{figure}

Comparison of Figures~\ref{fig:GenbodMult18LCMSK0p5},~\ref{fig:GenbodMult9LCMSK0p5} and~\ref{fig:GenbodMult6LCMSK0p5}
makes clear the multiplicity dependence of the EMCICs.  As expected, lower multiplicity
events show a greater effect.  Also (not shown), increasing the amount of energy to be
distributed among the particles, for fixed multiplicity, decreases EMCICs, as one expects.


Finally, we note that EMCICs can affect the correlation function even down to very low $Q$, again
reminding us that we cannot (responsibly) ignore these effects in a femtoscopic analysis.

\section{ANALYTIC CALCULATION OF EMCICs}
\label{sec:JYO}

Now then, EMCIC effects generated by \genbod ``resemble'' the experimental data, but it is likely unwise
to use \genbod itself to correct the data for several reasons.  Firstly, there is strong sensitivity to
the (not completely measured) number and species-mix of {\it all} particles emitted in the event, including
neutrinos and possible magnetic monopoles (or, less exotically, particles escaping detector acceptance).
Secondly, there is strong sensitivity to the energy ``available'' in the event; it is not obvious that this
is $\sqrt{s_{NN}}$ of the collision.  Clearly, EMCIC effects depend on the individual momenta $\vec{p}_1$ and
$\vec{p}_2$ of the particles entering the correlation function.  This will depend on acceptance, efficiency,
kinematic cuts (both purposeful and those imposed, e.g., by particle-identification limitations),
and, to a degree, the underlying single-particle phasespace.  (While correlation functions are insensitive to
the single-particle phasespace, the correlations which they measure may, in fact, depend on this phasespace,
due to physical effects.)

Thus, one would like to calculate EMCICs, based on the data itself.  An excellent start in this direction
has been presented by Danielewicz~\cite{Danielewicz:1987in} and later by Borghini, Dinh and Ollitraut~\cite{Borghini:2000cm},
in which they considered the effect of EMCICs on two-particle azimuthal correlations (elliptic flow $v_2$).
They considered transverse momentum ($\vec{P}_T$) conservation only, but the extension of their formula
to three, and even four~\footnote{The authors thank U. Heinz for showing that on-shell conditions do not cause major complications
               in the extension to the formula to include energy conservation.}
dimensions is straightforward.

We consider the single-particle distribution of a mass-$m_i$ particle {\it un}affected by EMCICs
\begin{equation}
\tilde{f}(p_i) \equiv 2E_i\frac{d^3N}{d\vec{p}_i^3} .
\end{equation}
The $k$-particle distribution ($k$ less than total multiplicity $N$), {\it including} EMCICs is then
\begin{eqnarray}
\label{eq:JYO1}
\tilde{f}_c\left(p_1,...,p_k\right) = \left(\prod_{i=1}^k \tilde{f}(p_i)\right)\times  \qquad \qquad \qquad \qquad\nonumber \\
\frac{\int\left(\prod_{j=k+1}^N d^4p_j \delta\left(p^2_j-m^2_j\right)\tilde{f}\left(p_j\right)\right)\delta^4\left(\sum_{i=1}^N p_i-P\right)}
{\int\left(\prod_{j=1}^N d^4p_j \delta\left(p^2_j-m^2_j\right)\tilde{f}\left(p_j\right)\right)\delta^4\left(\sum_{i=1}^N p_i-P\right)} .
\end{eqnarray}
(Note the difference between numerator and denominator in the starting value of the index $j$ on the product.)

According to Equation~\ref{eq:JYO1}, the $k$-body momentum distribution, including EMCICs, is the
$k$-body distribution {\it not} affected by EMCICs-- i.e. just an uncorrelated product of single-particle distributions--
multiplied by a ``correction factor'' which enforces the EMCIC.  The numerator of this factor just demands that the remaining $N-k$ on-shell
particles are configured so as to conserve total energy and momentum, and the denominator just normalizes the distribution.

Using the central-limit theorem
 (valid for ``large enough'' $N-k$), we find
\begin{eqnarray}
\label{eq:JYO2}
\tilde{f}_c\left(p_1,...,p_k\right) = \left(\prod_{i=1}^k \tilde{f}(p_i)\right)\times \qquad \qquad \nonumber \\
\left(\frac{N}{N-k}\right)^2\cdot \exp\left[-\sum_{\mu=0}^3\frac{\left(\sum_{i=1}^k\left(p_{i,\mu}-\langle p_{\mu} \rangle \right)\right)^2}
{2\left(N-k\right)\sigma_\mu^2}\right]
\end{eqnarray}
where
\begin{equation}
\label{eq:defineSigma}
\sigma_\mu^2 \equiv \langle p_\mu^2 \rangle - \langle p_\mu \rangle^2 
\end{equation}
and
\begin{equation}
\label{eq:defineRMS}
\langle p^2_\mu \rangle \equiv \int dp \tilde{f}(p) \cdot p_\mu^2 .
\end{equation}
Naturally, $\langle p_{(\mu=1,2,3)} \rangle = 0$.
(In these equations, we now assume only one species of particles, so that no species label is needed for $\langle p_\mu^2 \rangle$.
This is only for simplicity of notation here; results, including the ``experimentalist's formula'' below, only become more
cumbersome to write, but are similar otherwise.)

Note that even the single-particle momentum distribution is affected by EMCICs
\begin{eqnarray}
\tilde{f}_c\left(p_i\right) = \tilde{f}\left(p_i\right) \cdot \left(\frac{N}{N-1}\right)^2 \times \qquad \qquad \qquad \qquad \\
\exp\left[-\frac{1}{2(N-1)}\left(
\frac{p^2_{i,x}}{\langle p_x^2 \rangle}+\frac{p^2_{i,y}}{\langle p_y^2 \rangle}+\frac{p^2_{i,z}}{\langle p_z^2 \rangle}
+\frac{\left(E_i-\langle E \rangle\right)^2}{\langle E^2 \rangle -\langle E \rangle^2}\right)\right] \nonumber
\end{eqnarray}

The $k$-particle correlation function is defined as the measured (i.e. EMCIC-affected) $k$-particle yield divided by
the product of the measured single-particle yields
\begin{eqnarray}
\label{eq:JYOc2}
C\left(p_1,...,p_k\right)  \equiv 
  \frac{\tilde{f}_c\left(p_1,...,p_k\right)}{\tilde{f}_c\left(p_1\right)\cdots\tilde{f}_c\left(p_k\right)} 
   = 
  \frac{\left(\frac{N}{N-k}\right)^2}{\left(\frac{N}{N-1}\right)^{2k}} \times \qquad \\
%
%
%
  \frac{\exp\left[\frac{-1}{2(N-k)} \left\{
     \sum_{\mu=1}^{3}\left(\frac{\left(\sum_{i=1}^k p_{i,\mu}^2\right)^2}{\langle p_\mu^2\rangle}\right)
     +
     \frac{\left(\sum_{1}^k \left( E_i - \langle E\rangle\right)\right)^2}{\langle E^2 \rangle - \langle E \rangle^2}\right\}
      \right]}
     {\exp\left[\frac{-1}{2(N-1)}\sum^{k}_{i=1}\left\{
       \sum_{\mu=1}^3\frac{p_{i,\mu}^2}{\langle p_\mu^2\rangle}
       +
       \frac{\left( E_i - \langle E\rangle\right)^2}{\langle E^2 \rangle - \langle E \rangle^2}\right\}\right]}  \nonumber
\end{eqnarray}

An important point: EMCICs result from the constraint that the event's energy-momentum is the same fixed number for all pairs in the event.
This is true in the laboratory frame, but not in LCMS or pair rest frame.  Thus, while one may {\it bin} the
correlation function in the frame of one's choice, the momenta which appear on the right-hand-side of Equation~\ref{eq:JYOc2} must
be in the laboratory system.

To first order in $1/N$, the two-particle correlation function becomes
\begin{eqnarray}
\label{eq:JYOc2firstorder}
C(p_1,p_2)=1- \qquad \qquad \qquad \qquad \qquad \qquad \\
     \frac{1}{N}\left(
     2\frac{\vec{p}_{1,T}\cdot\vec{p}_{2,T}}{\langle p_T^2 \rangle} + 
      \frac{p_{1,z}\cdot p_2,z}{\langle p_z^2 \rangle} +
      \frac{\left(E_1-\langle E \rangle\right)\left(E_2-\langle E \rangle\right)}{\langle E^2\rangle-\langle E\rangle^2}\right) \nonumber
\end{eqnarray}
where we have taken $\langle p_x^2 \rangle = \langle p_y^2 \rangle = \langle p^2_T \rangle/2$.
In what follows, we shall refer to the first, second, and third terms within the parentheses of Equation~\ref{eq:JYOc2firstorder}
as the ``$p_T$ term,'' ``$p_z$ term,'' and ``$E$ term,'' respectively.

If we know $N$, $\langle p_T^2 \rangle$, $\langle p_z^2 \rangle$, $\langle E^2 \rangle$, and $\langle E \rangle$ from
the data, we can calculate EMCICs using Equation~\ref{eq:JYOc2}.
Better yet, if $N$ is large enough, then we can use Equation~\ref{eq:JYOc2firstorder}.
This is what is done in Figures~\ref{fig:GenbodMult18-noEtaCut}-\ref{fig:GenbodMult9LCMSK0p5}.
The black circles, blue stars, and red triangles 
show the $p_T$, $p_z$ and $E$ terms, respectively, from the first-order expansion (Equation~\ref{eq:JYOc2firstorder}), while
the open circles and orange inverted triangles represent the results of Equation~\ref{eq:JYOc2firstorder} and
Equation~\ref{eq:JYOc2}, respectively.

Several observations are in order.  
Firstly, each of the three terms in Equation~\ref{eq:JYOc2firstorder} produce non-trivial behaviour of the \Alm's,
interfering with each other in interesting ways.  We find also that the $p_z$ term affects $A_{2,2}$; this was initially
surprising since $A_{2,2}$ quantifies the behaviour of the correlation function in the ``out-side'' plane, while $\hat{z}$ is
the ``long'' direction in the Bertsch-Pratt system.\footnote{This raises the issue of whether conservation of $p_z$ and energy
are negligible for $v_2$ measurements, as has been assumed.}
Clearly, EMCICs projected onto a 2-particle space are non-trivial objects.

It is seen that the first-order expansion (Equation~\ref{eq:JYOc2firstorder}) agrees well
with the full expression (Equation~\ref{eq:JYOc2}) well for $N > \sim 10$.  Such multiplicities are relevant for the $p+p$
measurements done at RHIC (especially recalling that $N$ includes all particles, even unmeasured ones).  We see also that
the analytic calculations (open circles and inverted triangles) approximate the results of the \genbod simulation (green squares),
especially as the multiplicity and total energy of the event increases; increasing agreement for large $N$ and $E_{tot}$ is expected,
given the approximations leading to our analytic expressions.
We observe also that the analytically-calculated expressions respond identically to the kinematic cuts as does the simulation
(c.f. Figures~\ref{fig:GenbodMult18-noEtaCut} and ~\ref{fig:GenbodMult18-withEtaCut}).

Never do the analytic calculations never reproduce {\it exactly} the simulations; we discuss this further in the next Section.

\section{AN EXPERIMENTALIST'S FORMULA}
\label{sec:formula}

Even for large $N$ and energy, the calculations do not exactly reproduce the EMCIC effects in the simulation.
One reason for this may be found, in fact, in the definition of the average values (e.g. $\langle p_z^2 \rangle$) themselves.
In Equation~\ref{eq:defineRMS}, average quantities are calculated using the distribution $\tilde{f}(p)$, which is
not affected by EMCICs.  Naturally, the only measurable distribution available to the experimentalist (even when \genbod
simulations serve as the ``experiment'') is $\tilde{f}_c(p)$.

Thus, it appears the experimentalist cannot plug her data into the equations~\ref{eq:defineSigma}, \ref{eq:defineRMS}
and~\ref{eq:JYOc2firstorder} to fully calculate EMCICs.  However, such an ambition would have been hopeless anyhow.
After all, even the total multiplicity $N$ (again, including photons etc) is rarely fully measured.  And finite kinematic
acceptance (e.g. in $\eta$) will require extrapolation to calculate, e.g. $\langle p_z^2 \rangle$.

To the practicing femtoscopist, there is a natural solution.
Having at hand (1) educated guesses for the quantities $N$, $\langle E^2 \rangle$ etc, and (2) a physically-motivated
functional form which connects these quantities to the correlations we'd like to understand, we perform a fit.
Let us rewrite Equation~\ref{eq:JYOc2firstorder} as
\begin{eqnarray}
\label{eq:formula}
C\left(p_1,p_2\right) = 1
   - M_1\cdot\overline{\left\{\vec{p}_{1,T}\cdot\vec{p}_{2,T}\right\}}
   - M_2\cdot\overline{\left\{p_{1,z}\cdot p_{2,z}\right\}} \quad \\
   - M_3\cdot\overline{\left\{E_1 \cdot E_2\right\}}
   + M_4\cdot\overline{\left\{E_1 + E_2\right\}}
   - \frac{M_4^2}{M_3} . \qquad \nonumber
\end{eqnarray}
where
\begin{eqnarray}
\label{eq:fitParameters}
M_1\equiv\frac{2}{N\langle p_T^2 \rangle}
  \quad , \quad  M_2\equiv \frac{1}{N\langle p_z^2 \rangle} \qquad \qquad \qquad \nonumber \\
M_3\equiv\frac{1}{N\left(\langle E^2\rangle-\langle E\rangle^2\right)}  
  \quad , \quad M_4\equiv\frac{\langle E\rangle}{N\left(\langle E^2\rangle-\langle E\rangle^2\right)} .
\end{eqnarray}

The notation $\overline{\left\{X\right\}}$ in Equation~\ref{eq:formula} highlights the fact that $X$ is
a two-particle quantity which depends on $p_1$ and $p_2$ (or $\vec{q}$, etc): $\overline{\left\{X\right\}}\left(\vec{q}\right)$.
From a practical point of view, $X$ is averaged over the same $\vec{q}$ bins as used for the correlation function.  This
involves nothing more than adding four more histograms to the several already being constructed by the experimentalist
as she runs through all pairs in the data.   
The binned functions $\overline{\left\{X\right\}}$ then automatically reflect the same event and particle selection 
as the correlation function.

It is appropriate here to re-emphasize the point made in reference to Equation~\ref{eq:JYOc2}.  The event's total energy
and momentum is a fixed quantity in a fixed (e.g. lab) frame; in particular, the momentum in Equation~\ref{eq:JYOc2} is
assumed $\vec{P}=\vec{0}$-- i.e. the collision-center-of-mass (CCM) frame is assumed.
In a pair-dependent frame (e.g. pair center-of-mass PCM or longitudinally co-moving system LCMS),
the event's energy and momentum will depend on the pair.  EMCICs, therefore, must be calculated with CCM momentum.
Thus, in the function $\overline{\left\{p_{1,z}\cdot p_{2,z}\right\}}\left(\vec{q}\right)$, $p_{1,z}$ and $p_{2,z}$ must be calculated in
the CCM frame, while the {\it binning variable} $\vec{q}$ should be in whatever frame one chooses to construct the correlation function $C$.

The parameters $M_i$ defined in Equation~\ref{eq:fitParameters}, on the other hand, are global and independent
of $p_1$ and $p_2$.  It is these which we will use as fit parameters.  The task is then fast and straightforward;
the EMCIC part of the correlation function $C(\vec{q})$ is simply a weighted sum of four functions.  Indeed, 
one may calculate coefficients as in Equation~\ref{eq:Alm} for the four new functions.  For example
\begin{eqnarray}
A_{l,m}^{p_Z}\left(Q\right) \equiv 
\sum_{{\rm bins} \thinspace i} \overline{\left\{p_{1,z}\cdot p_{2,z}\right\}}\left(Q,\cos\theta_i,\phi_i \right) \cdot
\qquad\qquad \\
Y_{l,m}\left(\cos\theta_i,\phi_i\right) 
    F_{l,m}\left(\cos\theta_i,\Delta_{\cos\theta},\Delta_\phi\right) , \nonumber
\end{eqnarray}
etc.
Then, thanks to the linearity of Equation~\ref{eq:formula} and the orthonormality of $Y_{l,m}$'s, the measured \Alm's themselves are similarly
just weighted sums of harmonics
\begin{eqnarray}
\label{eq:AlmFit}
\AlmQ = \delta_{l,0}\cdot\left(1-M_4^2/M_3\right) 
  - M_1\cdot A_{l,m}^{p_T}\left(Q\right)             \qquad \\
  - M_2\cdot A_{l,m}^{p_Z}\left(Q\right)
  - M_3\cdot A_{l,m}^{(E\cdot E)}\left(Q\right)
  + M_4\cdot A_{l,m}^{(E+E)}\left(Q\right) . \nonumber
\end{eqnarray}
Treating Equation~\ref{eq:AlmFit} as a fit, we have a few (say six, for $l\leq 4$) {\it one}-dimensional functions to fit with four adjustable weights.

\begin{figure}[t!]
{\centerline{\includegraphics[width=0.5\textwidth]{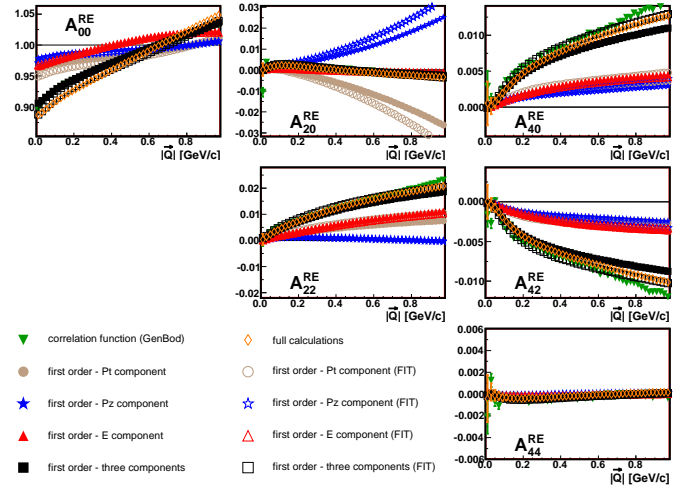}}}
\caption{\Alm's from 18-pion \genbod-generated events.
Green inverted triangles (often underneath black squares) is the correlation function (measured in PRF) from \genbod.
Filled brown circles, filled blue stars and filled red triangles show, respectively, the ``$p_T$,'' ``$p_z$,'' and ``$E$'' terms, defined in Equation~\ref{eq:JYOc2firstorder};
black filled squares show their sum.
Open symbols of the same shape and color (identified as ``FIT'' in the legend) show corresponding terms, except with weights (see Equation~\ref{eq:fitParameters})
adjusted to maximize agreement between the open black squares and the simulation.\label{fig:FitAlmGenbod}}
\end{figure}

A first example of such a fit is shown in Figure~\ref{fig:FitAlmGenbod}.  Again the \genbod simulation is compared to the first-order form
of Equation~\ref{eq:formula}.  The filled circles, stars and triangles show the ``$p_T$'' ($M_1$), ``$p_z$'' ($M_2$), and ``E'' ($M_3$ and $M_4$) terms
when the weights (Equation~\ref{eq:fitParameters}) are calculated directly from the events, as discussed in Section~\ref{sec:JYO}.  Treating the $M_i$
as adjustable parameters leads to a slightly different weighting of the terms, and a slightly better fit to the data.

Lest we forget, our original goal was not to understand EMCICs per se, but to extract the femtoscopic information from measured two-particle correlations.
Assuming that the only non-femtoscopic correlations are EMCICs, one may simply add the femtoscopic terms (e.g. Gaussian in $(q_o,q_s,q_l)$ space or whatever)
to the fitting function in Equation~\ref{eq:formula} or~\ref{eq:AlmFit}.  Common femtoscopic fitting functions usually contain $\sim 5$ parameters (e.g. $R_{out}$)
In the imaging technique~\cite{Brown:1997ku},
one assumes the separation distribution is described by a sum of splines (rather than, say, a Gaussian); here, too, there are usually
4-5 fit parameters (spline weights).
We have found that the number of fit parameters now must be doubled to account also for EMCICs.  
This is a non-trivial increase in analysis complexity.  However, we keep in mind two points.

Firstly, the increased work is neccessary.  EMCICs (and possibly other important non-femtoscopic correlations) are present and increasingly relevant at low multiplicity.
One option is to ignore them, as has sometimes been done in early high-energy experiments.  However, with the new high-quality data and desire for high-detail understanding at RHIC,
ignoring obvious features such as those seen in Figures~\ref{fig:CartesianDAu} and~\ref{fig:SHDdAu} is clearly unacceptable.
Perhaps a slightly better option is to invent an ad-hoc functional form with no real physical basis (and often
manifestly wrong symmetry~\citep[cf][]{Chajecki:2006sf}), which introduces new parameters in any case.  We hope that the results here
present a relatively painless, and considerably more responsible, third option.

Secondly, while the non-femtoscopic EMCICs are not confined to the large-$Q$ region (an important point!), the femtoscopic correlations are confined
to the small-$Q$ region.  Therefore, one hopes that the addition of four new parameters to the fit of the correlation function will not render the
fit overly unwieldy.  While we can not expect complete block-diagonalization of the fit covariance matrix, one hopes that the $M_i$ are determined
well enough at high $Q$ that the femtoscopic fit parameters can be extracted at low $Q$.

\section{SUMMARY}
\label{sec:summary}

To truly claim an understanding of the bulk nature of matter at RHIC and the LHC, a detailed picture of the
dynamically-generated geometric substructure of the system created in heavy ion collisions
is needed.  It is believed that this substructure,
and the matter itself, is dominated by strong collective flow.  The most direct measure of this flow is
a measurement of the space-momentum correlation (e.g. $R\left(m_T\right)$) it generates.  The physics
of this large system, and the signals it generates, should be compared to the physics dominating $p+p$
collisions, as is increasingly common in high-$p_T$ studies at RHIC.  
For the small systems, however,
non-femtoscopic effects contribute significantly to the correlation funcion, clouding the extraction
and interpretation of the femtoscopic ones.  

EMCICs, correlations generated by kinematic conservation laws, are surely present and increasingly relevant
as the event multiplicity is reduced.  Using the code \genbod to study correlation functions solely driven
by EMCICs, we found highly non-trivial structures strongly influenced by event characteristics (multiplicity
and energy) and kinematic particle selection.

We extended the work of Danielewicz and Ollitrault to include
four-momentum conservation and applied it to correlation functions commonly used in femtoscopy.  We found
structures associated individually with the conservation of the four-momentum components, which interfere
in nontrivial ways.  
Comparison of the analytic EMCIC calculations with the \genbod simulation gave confidence that
the approximations (e.g. ``large'' multiplicity $N$) entering into the calculation were sufficiently
valid, at least for multiplicities considered here.
We further showed that the full EMCIC calculation can safely be replaced
with a first-order expansion in $1/N$.

Based on this first-order expansion, we developed a practical,
straight-forward ``experimentalist's formula''
to generate histograms from the data which are later used in a generalized fit to the
measured correlation function, including EMCICs and femtoscopic correlations.

The huge systematics of results and interest in femtoscopy in heavy ion collisions is
renewing similar interest in the space-time signals from $p+p$ collisions.  Direct comparisons
between the two systems are now possible at RHIC and have already produced intriguing (albeit preliminary)
results.  Very soon, $p+p$ collisions will be measured in the LHC experiments, and the heavy ion experimentalists
will be eager to apply their tools.  The femtoscopic tool is one of the best in the box-- so long as
we keep it sufficiently calibrated with respect to non-femtoscopic effects increasingly relevant in small systems.

\medskip

We would like to thank the organizers of this workshop-- most especially the tireless Dr. Sandra Padula--
for arranging a enjoyable gathering of experts in a very productive environment.  We wish to thank Drs. 
Ulrich Heinz, Adam Kisiel, Konstantin Mikhaylov, Jean-Yves Ollitrault, 
and Alexey Stavinsky for important suggestions and insightful discussions.

\bibliographystyle{annrev}
\bibliography{SmallSystems}

\end{document}